\documentclass[pre,twocolumn,showpacs]{revtex4}
\usepackage{epsf}
\usepackage{epsfig}
\begin{document}
\title{
L\'evy Statistics for Random Single--Molecule Line Shapes in a Glass
}
\author{E. Barkai$^a$, A. V. Naumov$^b$, Yu. G. Vainer$^b$, M. Bauer$^c$, L. Kador$^c$ \\
$^a$ Department of Chemistry and Biochemistry, Notre Dame University, Notre Dame, IN 46556.\\
$^b$
Institute of Spectroscopy, Russian Academy of Sciences, Moscow Reg., Troisk
142190, Russia\\
$^c$
Institute of Physics and BIMF, University of Bayreuth, D-95440 Bayreuth, Germany\\
}

\date{\today}

\begin{abstract}

 We demonstrate that the statistical behavior of random line shapes of
single tetra-tert-butylterrylene chromophores embedded 
in an amorphous polyisobutylene matrix at $T=2$ K,
is described by L\'evy statistics as predicted theoretically
by Barkai, Silbey, and Zumofen
[{\em Phys. Rev. Lett.}
{\bf 84} 5339 (2000)]. This behavior is
a manifestation of the long range interaction between two level
systems in the glass and the single molecule. A universal
amplitude ratio is investigated which shows that the standard
tunneling model assumptions are compatible with the experimental data.

\end{abstract}

\pacs{61.43.Fs, 05.40.Fb, 78.66.Jg}

\maketitle

 Experimental advances \cite{MO} have made it possible to measure
the fluorescence of a single molecule (SM)
embedded in a glass.
Because each individual molecule is in a unique static and dynamic environment,
the fluorescence of chemically identical SMs varies from molecule
to molecule. 
In this way the molecules serve as local reporters
on the dynamics and statics of the host glass. 
SM experiments have been performed both in low-temperature
glasses 
\cite{Zumb1,Zumb2,Orrit,Elisa,Donley2}
and recently close to the glass transition
temperature 
\cite{Desch1}. For low-temperature
glasses the fundamental question is: 
Is the standard
tunneling model valid for glasses?
Related questions are how to analyze the
complex line shape behaviors of SMs
in glasses, and what do their random behaviors 
teach us on the SM-glass system.

The standard tunneling model
\cite{AHV} 
was
suggested in the early seventies to explain universal features
of glasses; for example many glasses
show a heat capacity which is nearly linear in temperature. 
At the center of this phenomenological
model is the concept of the two-level system (TLS).
It is assumed that at cryogenic temperatures excitations in
glassy materials are two-level tunneling systems
whose
energies and tunneling matrix elements are randomly distributed. 
More recently, 
Geva and Skinner  \cite{Geva} modeled behaviors of SMs
in glasses based on the standard tunneling model.
Orrit and co-workers \cite{Zumb1,Zumb2}
used the fluorescence of single terrylene molecules in 
the polymer polyethylene
to obtain the first {\em direct} experimental proof  
that two level systems  actually exist in
an amorphous material. 
More recent experiments \cite{Orrit}
revealed
behaviors not compatible with the
standard model for
$21$ out  of $70$ molecules;   
e.g., a molecule coupled to a three level system \cite{REM}.
We note that a fundamental first-principle understanding 
of TLSs is still missing \cite{Wol}, 
although numerical simulations \cite{HSilbey}
give some evidence on the microscopic nature of a few
of these entities.

 There are many open questions concerning 
the standard model. For example: Are the TLSs 
uniformly distributed in space, or do they preferentially
 appear at
boundaries of clusters of atoms/molecules, as suggested in \cite{Philips}.
Another open question is the nature of the interaction between
the TLSs and the SM e.g., is it dipolar as suggested
in \cite{Geva}? A method to obtain this important information using SM
spectroscopy was suggested in
\cite{Barkai} (see details below). 

  A theoretical investigation of
the distribution of random line shapes of SMs in
glass was carried out in \cite{Barkai,BZS} based on the
standard model approach
\cite{Geva}.
 Interestingly, the theoretical results
obtained in \cite{Barkai}
  showed that 
the problem of random line shapes of SMs 
in glasses is related to L\'evy statistics,
thus the generalized
Central Limit Theorem \cite{Feller} applies to this problem. 
This connection to L\'evy statistics is a manifestation of long--range
interactions between the TLSs in the glass and the SM
(see details below).
We note that L\'evy statistics is known to describe several
other long-range interaction models in diverse fields
such as turbulence \cite{Min},  
and
random magnetic systems \cite{Maass}.
Stoneham's theory \cite{Stoneham} of inhomogeneous
line broadening in crystals with defects is based on long-range forces, and 
it can be interpreted in terms of L\'evy
statistics \cite{BZS}.

\begin{figure}[htb]
\epsfxsize=20\baselineskip
\centerline{\vbox{
        \epsffile{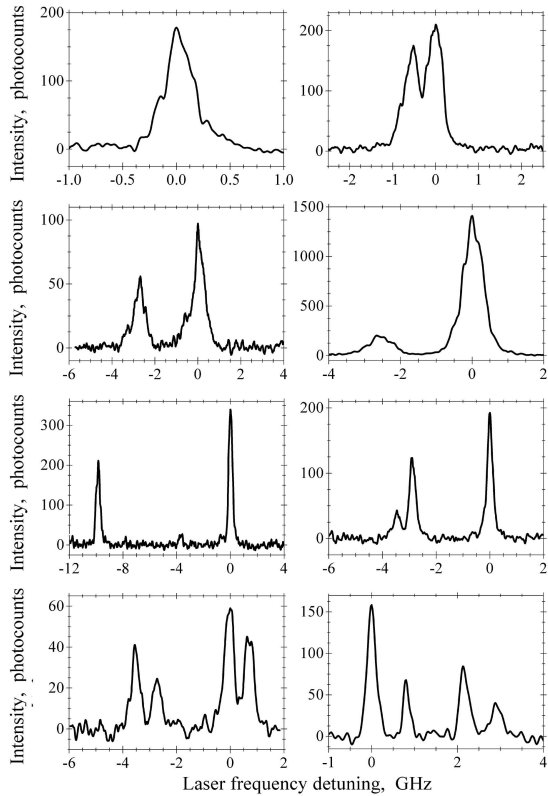}  }}
\caption {
Line shapes 
of 
single tetra-tert-butylterrylene chromophores embedded 
in an amorphous polyisobutylene matrix at $T=2$ K.
Note the doublet
and quartet features of some of the line shapes 
due to strong coupling to one or two TLSs. One of the molecules
has three peaks, indicating the possibility that this molecule is coupled to
a three--level system. This {\em rare} type of behavior is not consistent
with the standard tunneling model. 
Another possibility
is that measurement time
was not long enough in this case.
}
\label{fig0a}
\end{figure}

 In this paper we analyze the statistical
properties of random line shapes in a glass
and compare it to the theoretical predictions in \cite{Barkai}.
In Fig. \ref{fig0a} we show eight lines of
single tetra-tert-butylterrylene chromophores embedded 
in an amorphous polyisobutylene matrix at $T=2$ K. 
For experimental details see 
\cite{Naumov}.
The lines are typically multi-peaked similar to the
numerical predictions in  \cite{Geva}.
Each line is different from any other line, since each individual
SM is in a unique environment.
The multi-peaked
behavior of the line shapes in Fig. \ref{fig0a} can be 
qualitatively explained using standard model arguments.
If a SM
is coupled to a single slow flipping TLS embedded in
its vicinity, we expect 
that when the TLS flips from its up state to its down state or vice versa,
the SM absorption frequency will shift.
In this case the line shape of the SM
is a doublet.
It follows that the frequency of a SM coupled to $N$ such independent
two-level
systems will jump between $2^N$ states.
Hence as shown in Fig. \ref{fig0a},
the single molecule's line shape will be composed of
$2^N$ peaks. The width of these peaks depends on the dynamics
of TLSs situated far from the molecule, fast dephasing processes,
and the life time of the electronic transition 
\cite{Geva,Barkai,ADV,Brown}.  

 To obtain the lines in Fig. \ref{fig0a} we used the spectral
trail technique introduced by Moerner and
co-workers \cite{MT}, following the spectral activity
of the molecule during a scan time which in our experiments
was fixed to be $120$ Sec.
Following the jump history of the SM  enables us to identify
the peaks in Fig. \ref{fig0a} as originating from a SM.
Without using this method it is practically
impossible to say if the lines in Fig. \ref{fig0a} are due to
contribution from several SMs or originate from a SM. 

 Random multi-peaked lines as shown in Fig. \ref{fig0a} are
a novel feature of SM spectroscopy, not observed
in ensemble averaging techniques. 
To obtain information on the SM-glass system we
investigate the distribution of the line shapes of
SMs in a glass. Mathematically we are dealing with the
question of the distribution function of a function
[i.e., the line $I(\omega_L)$ is a function of
laser frequency].
 Recently, Refs. \cite{Barkai,BZS},
suggested to characterize the
line shape of each molecule by its cumulants $\kappa_1,\kappa_2,\kappa_3,...$.
Thus each line is characterized by an infinite set of cumulants,
which are random variables that vary from one
molecule to the other, thus
 reflecting the disordered nature of the
glass.
 The distribution functions $P(\kappa_1),
P(\kappa_2),P(\kappa_3),\cdots$  completely characterizes the statistical
properties of 
the line shapes of SMs in a glass. 
 The cumulants are obtained from the
moments of the line shape \cite{Naumov2},
$m_n = \int \omega_L ^n I \left( \omega_L  \right) {\rm d} \omega_L $,
according to the well-known relations 
$ \kappa_1 = m_1$,
$\kappa_2 = m_2 - (m_1)^2$ etc.
From $244$ SM spectra, we obtained the histograms
of the first two cumulants,
$P(\kappa_1),P(\kappa_2),$ which 
yield important information on the glass--SM
system. 
We consider here the distribution of the cumulants and not the
distribution of the moments, since the cumulants were predicted
to be described by L\'evy statistics \cite{Barkai}.

To better understand the meaning of our data analysis
and the relationship to L\'evy statistics,
it is useful to recall five main assumptions of the
model used in \cite{Geva,Barkai}. \\
(i) The absorption frequency of the SM
follows a stochastic trajectory described by 
$\omega(t) = \omega_0 + \sum_{n=1}^N \xi_n(t) \nu_n(r)$,
where $\omega_0$ is the bare frequency of the molecule. 
$\xi_n(t)$ are random functions of time which follow
a two-state process, $\xi_n(t)=1$ when the $n$-th TLS is
in state up or $\xi_n(t)= 0$ when it is in state down.
Thus the flips of the TLSs induce spectral diffusion.
 The flipping rate between the up and down states is determined
by a rate $R$ which varies from TLS to TLS [the distribution
of jumping rates of TLSs in a glass spans many orders of magnitude
from nano seconds to (at least) days]. 
(ii)  The SM frequency shifts are:
$\nu_n = 2 \pi \alpha \Psi\left( \Omega \right) {A\over E } {1 \over  (r_n)^3}$.
The most important ingredient of the theory is
the long-range interaction between the TLS and
SM, $\nu \propto 1/ r^3$, reflecting the assumption
of dipolar interaction. 
This long-range type of interaction
is the first important ingredient in the relationship
between SM spectroscopy in glasses and L\'evy statistics.
Other parameters controlling the frequency shifts are:
$\alpha$ the SM-TLS coupling constant, 
$E=\sqrt{A^2 + J^2}$ the energy splitting of the TLS,
$\Psi\left( \Omega \right)$ a dimensionless function
of order unity
describing the
orientation of the TLS and SM, and finally 
the random parameters of the TLS: $A$ (asymmetry parameter)
 and $J$ (tunneling matrix element).
(iii) The TLSs are uniformly distributed is space and are non-interacting.
This assumption is the second important condition for L\'evy 
statistics to be valid.
(iv) The standard tunneling
model is valid; this model determines the distribution
of parameters describing $\nu_n$ and $R_n$
as well as the density of TLSs. 
Note, however, that the L\'evy statistics and results in \cite{Barkai}
are not limited to 
this model. 
(v) The stochastic Kubo--Anderson theory 
of line shapes
is applicable, implying 
weak laser fields.
Under these conditions the following two limiting 
behaviors were found \cite{Barkai}. 

 The first case corresponds
to the fast--modulation limit $\nu_n\ll R_n$
for all TLSs in the vicinity of the molecule. 
In this case, also called motional narrowing limit,
the
lines of individual molecules
are
Lorentzian in shape.
Then the lines
are characterized by two parameters only, e.g., the
width at half maximum and the  center location. 
The distributions of these two parameters
are L\'evy stable laws \cite{Barkai}.
From Fig. \ref{fig0a} it is clear that
the fast modulation limit does not describe our
experimental results.

\begin{figure}[htb]
\epsfxsize=20\baselineskip
\centerline{\vbox{
        \epsfig{file=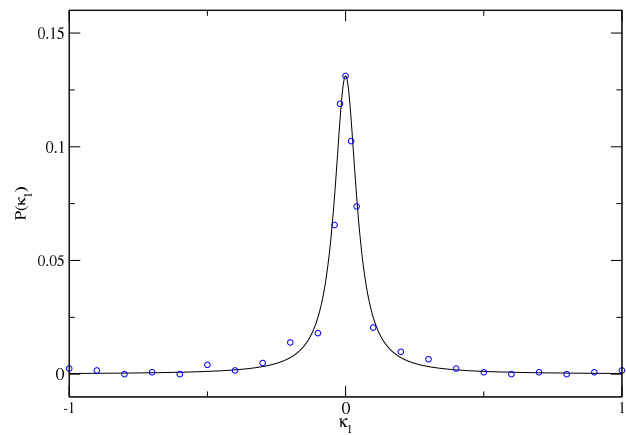, width=0.85\linewidth, angle=-90}  }}
\caption {
Probability density of
the first cumulant $\kappa_1$ (units are GHz). 
The dots are experimental
results, the curve is a one-parameter fit to a Lorentzian.
}
\label{fig1}
\end{figure}

 The second case corresponds to  
the slow-modulation limit. 
If $R_n \ll \nu_n$ for all the TLSs in the vicinity of the molecule, 
the {\em shape} of the line is random
and typically multi--peaked \cite{Barkai}. 
In this slow--modulation limit 
the distributions of line shape cumulants, 
$P(\kappa_1),P(\kappa_2),P(\kappa_3),$
etc,
are all L\'evy stable.
 Specifically,
the probability density function of the first cumulant $\kappa_1$
is given by
the symmetrical L\'evy density, $P(\kappa_1)= l_{1,0}(\kappa_1)$,
namely the Lorentzian
\begin{equation}
P(\kappa_1)={1 \over \pi} {z_1 \over \kappa_1^2 + z_1^2}, 
\label{eqLor}
\end{equation}
where $z_1$ is a scaling parameter which can be calculated from
the theory in \cite{Barkai}.
In Fig. \ref{fig1} we show that our experimental results
are compatible with the theoretical prediction.
We also fitted our results to a Gaussian probability density (not shown)
and found that the distribution of the first cumulant is definitely
not Gaussian. 

 We note that the reference frequency,
determining the laser detuning in  Fig. \ref{fig0a},
 was choosen on the maximum intensity of
the spectrum of the SM  $\omega_{max}$.
 While the theoretical
reference frequency in \cite{Barkai}
was the bare frequency of the molecule $\omega_0$.
Using numerical simulation based on the approach
in \cite{Geva}, we observed
that also when the reference frequency is choosen as $\omega_{max}$
distribution of first cumulant is well fitted by a Lorentzian,
in agreement with the experimental results in Fig.
\ref{fig1}. The value of $z_{1}$ in the two
approaches slightly differs, as we discuss below. We now discuss
the distribution of the second cumulant (i.e. the variance)
which is not sensitive
to definition of the reference frequency.

\begin{figure}[htb]
\epsfxsize=20\baselineskip
\centerline{\vbox{
        \epsfig{file=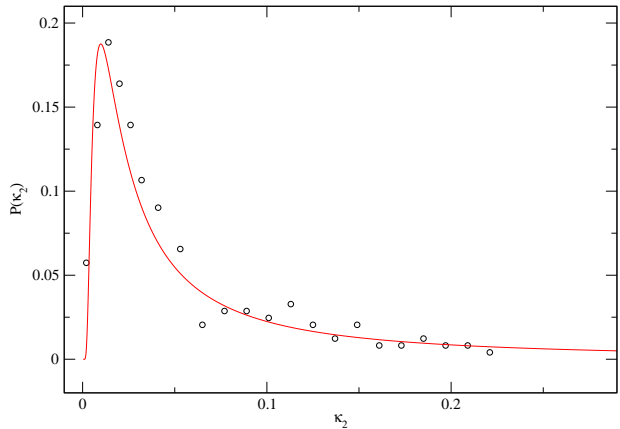, width=0.85\linewidth, angle=-90}  }}
\caption {
Probability density of
the second cumulant $\kappa_2$ (units $\mbox{GHz}^2$).
 The dots are experimental
results, the curve is a one-parameter fit to Smirnov's
probability density.
}
\label{fig4}
\end{figure}

The distribution of the second cumulant $\kappa_2$ 
is given by  the one-sided L\'evy stable law,
$P(\kappa_2)= l_{1/2,1}(\kappa_2)$,
namely 
 Smirnov's one-sided probability density 
\begin{equation}
P(\kappa_2) =  
{1 \over \left( z_{1/2}\right)^2}{2  \over \sqrt{\pi} } 
\left(  {2 \kappa_2\over z_{1/2}^2 } \right)^{-3/2} 
\exp\left( - {z_{1/2}^2\over 2 \kappa_2} \right),
\end{equation}
where the scaling parameter $z_{1/2}$ was derived  in \cite{Barkai}.
As shown in Fig. \ref{fig4},
the experimental data for the distribution of $\kappa_2$
are compatible with 
the theoretical prediction; the long tail of the
L\'evy stable law is visible.  
Yet, data of a larger number of molecules is needed
to improve the statistical fluctuations. The analysis of a larger number
of molecules will also enable us to compare theory and experiments
for the higher-order cumulants $\kappa_3$, $\kappa_4$ etc.  

 As mentioned, the L\'evy behavior is due to the
long-range dipolar interaction of  a SM with many TLSs,
hence the Gauss or L\'evy Central Limit Theorem arguments 
are expected to hold.
The L\'evy central limit theorem applies since
averaged frequency shifts diverge; 
$\langle \nu \rangle \propto \int_0^{\infty}
 r^{d-1}/r^3 {\rm d} r = \infty$ where $d=3$ is the dimensionality 
of the problem
(for mathematical details see \cite{Barkai}).
Hence, the L\'evy behavior obtained for the distributions
$P(\kappa_1)$ and $P(\kappa_2)$ is used to test the
assumptions of dipolar interactions and uniform-distribution
of the TLSs in space. The information about
the random distribution of the parameters of the glass,
i.e. $A$ and $J$,
are contained in 
the values of $z_{1/2}$ and $z_1$ which we now discuss.

%

 There are two distinct types of parameters describing
the SM-glass system: Those describing the bath of TLSs 
and the coupling constant $\alpha$ which depends on the
properties of the SM probe.
An important unsolved problem is how to extract information
from SM experiments which is sensitive only to the statistical
properties of the TLSs and is not affected
by $\alpha$ \cite{Elisa}.
The scaling parameters $z_1$ and $z_{1/2}$ depend on
$\alpha$,
the magnitude of these
two parameters depends also
on the precise modeling of the orientation function $\Psi(\Omega)$
 \cite{Barkai}.
Thus $z_1$ and $z_{1/2}$ are not universal functions, in the sense
that they depend on properties of the SM under investigation
and not on the properties of the glassy state
(which are supposed to be universal,
according to standard tunneling model). In fact, Donley et al \cite{Elisa}
suggested that
the coupling constant $\alpha$ itself should 
be a random variable.
This may seem to limit our ability to investigate 
low-temperature glasses with SM spectroscopy. 
If the statistical analysis of line shapes depends on an additional
unknown distribution function of the coupling constant (besides the 
standard distributions of the glass parameters), fitting of data to
the theory becomes rather arbitrary.  

 However, based on Eq. 9 in \cite{Barkai} 
one can show
that the ratio $z_{1/2}/z_1$ depends only on the statistical
properties
of the glass and not on the distribution
of the coupling constant $\alpha$. More precisely,
\begin{equation}
{z_{1/2} \over z_1} = { 1 \over \sqrt{ 2 \pi} } 
{\langle { A \over E} 
\mbox{Sech}\left( {E \over 2  k_b T } \right) \rangle_{AJ} 
\over \langle {A \over E} {1 \over 1 + \exp\left( E/ k_b T \right) }  \rangle_{AJ}},
\label{eqratio}
\end{equation}
where the averaging is performed  over
the TLS parameters $A$ and $J$  \cite{Barkai}. 
Since Eq. (\ref{eqratio}) is independent of the
exact distribution of $\alpha$, it is a useful tool
for describing the behavior of glasses. 
To derive Eq. (\ref{eqratio}) we assumed that the
random variable $\alpha$
is independent of the glass parameters $A$ and $J$.

%
 From our fits 
we find $z_{1/2} = 0.175$ GHz and $z_1= 0.0485$ GHz,
which yields $z_{1/2}/z_1=3.6$. The theoretical prediction,
based on Eq. (\ref{eqratio}) yields $z_{1/2}/z_1=2.4$. 
The deviation between theory and experiment is now explained.
As mentioned, the theoretical prediction is based on
the assumption that the bare frequency of the molecule $\omega_0$
is the reference frequency for measurement.
Using numerical simulations \cite{Geva} with the parameter set
relevant for our experiment (details will be published)
we find that the ratio of $z_1$ when 
$\omega_{0}$ is the reference frequency and $z_1$ when
$\omega_{\max}$ is the reference frequency, is
 $z_1 ^{\omega_0}/z_1 ^{\omega_{max}}=1.6$. Varying the value of 
$\alpha$ in our simulations, in the interval 
$10\mbox{GHz} \mbox{nm}^3 < \alpha< 40 \mbox{GHz} \mbox{nm}^3$ 
we observed that the ratio
 $z_1 ^{\omega_0}/z_1 ^{\omega_{max}}$ does
not depend on $\alpha$.
As expected $z_1 ^{\omega_0} > z_1 ^{\omega_{max}}$ since the value of
$\kappa_1$ becomes smaller (in statistical sense)
if we assign the origin to the
maximum of the spectrum. Using the correction
factor $z_1 ^{\omega_0}/z_1 ^{\omega_{max}}=1.6$
 we find that the theory yields $z_{1/2}/z_1=3.8$.
Taking into account that the standard tunneling model does not address 
the chemical composition of the disordered system or the
chemical and geometric details of the SM
under investigation, we believe 
that the theoretical result is in surprisingly
reasonable agreement
with experiment.  
Measurements of the ratio
$z_{1/2}/z_1$ for other types of SMs and glasses
and for wider range of control parameters
(i.e., temperature and scan time) 
will show whether SM
data are compatible with the
universal predictions of the standard model. 

 Our approach  based on the 
analysis of the distributions
of line shape cumulants is different from a second
approach used in other studies \cite{Geva,Zumb2,Elisa}.
Previously, the line was characterized by its width 
at half maximum and the distribution of the random line shapes
was described by distribution of line widths. 
Clearly such a method does not capture the multi--peaked
behavior of the SM lines shown in Fig.
\ref{fig0a}. An approach based on the distribution of line
widths is useful in the fast-modulation limit, which
as mentioned, (i) obeys L\'evy statistics but (ii) is
not relevant to the experiment under consideration.

 To summarize SM spectroscopy is an excellent method to
investigate disordered systems by removing the ensemble averaging
found in other techniques. We have shown that our experimental
results are compatible with L\'evy statistics
and with standard tunneling model predictions. 
In particular 
the following two assumptions are reasonable:
(i) The two-level systems are uniformly distributed in space,
(ii) The frequency shifts are caused by dipolar interactions
$\nu\propto {1 \over r^3}$. 
 We introduced the universal ratio, $z_{1/2}/z_{1}$
which is sensitive to details of the standard model,
but not to the coupling of the SM to the TLSs in the glass (i.e., not to
$\alpha$).
The comparison between the theoretical and
the experimental value of  this ratio can be used to test the validity of 
the standard model predictions.


\begin{thebibliography}{99}
\bibitem{REM} A similar analysis in our system shows that
the relative number of spectral trails
which are not consistent with the standard TLS model, was less than
a few percent.
\end{thebibliography}
\end{document}